\documentclass{article}

\usepackage{arxivstyle}

\usepackage{fontspec}
\usepackage{amsmath, amssymb, amsthm}
\usepackage{booktabs}
\usepackage{array}
\usepackage{multirow}
\usepackage{graphicx}
\usepackage{float}
\usepackage{caption}
\usepackage{placeins}
\usepackage{fancyvrb}
\usepackage{needspace}
\newsavebox{\promptboxbox}
\newenvironment{promptbox}{%
  \par\smallskip\noindent
  \begin{lrbox}{\promptboxbox}%
  \begin{minipage}{\dimexpr\linewidth-2\fboxsep-2\fboxrule\relax}%
  \setlength{\parindent}{0pt}%
}{%
  \end{minipage}%
  \end{lrbox}%
  \fbox{\usebox{\promptboxbox}}%
  \par\smallskip
}
\newcommand{\promptfont}{\ttfamily}
\newcommand{\prompthead}[1]{%
  \par\Needspace{0.55\textheight}%
  \smallskip\noindent\textbf{#1}\par\nobreak\smallskip
}
\newtheorem{theorem}{Theorem}
\usepackage[hidelinks]{hyperref}
\usepackage{url}
\usepackage{xcolor}
\definecolor{codegray}{gray}{0.95}

\usepackage{xeCJK}
\setCJKmonofont{FandolFang-Regular.otf}
\xeCJKsetup{CJKecglue=}

\graphicspath{{./}{figs/}{figs/fig1_motivation/}{figs/fig2_skill_token_dist/}{figs/fig3_pipeline/}{figs/fig_cluster_quality/}{figs/fig_cluster_projection/}{figs/fig_annotation_flow/}{figs/fig_bilateral_balance/}{figs/fig_taxonomy_distribution/}}

\newcolumntype{C}{w{c}{2.8em}}

\title{Skill Is Not Document: Query-Conditioned Compatibility for LLM Agent Skill Routing}

\author{
  Zifei Wang$^{1,*}$,\ \ Wei Wen$^{2,*\dagger}$,\ \ Qiang Ji$^{1}$,\ \ Keyu Chen$^{2}$,\ \ Ruizhi Qiao$^{1,2}$,\ \ Xing Sun$^{1,2}$ \\[2pt]
  $^{1}$Tencent IMA Product Center \quad $^{2}$Tencent Youtu Lab \\[2pt]
  {\small\texttt{\{zifeiwang, jawnrwen, peasirji, keyuchen, ruizhiqiao, winfredsun\}@tencent.com}}
}

\begin{document}
\maketitle
\lhead{}\chead{}\rhead{}
\pagestyle{plain}\thispagestyle{plain}
\renewcommand{\thefootnote}{\fnsymbol{footnote}}
\footnotetext[1]{Equal contribution.}
\footnotetext[2]{Corresponding author. Contact: \texttt{jawnrwen@tencent.com}.}
\renewcommand{\thefootnote}{\arabic{footnote}}
\setcounter{footnote}{0}

\begin{abstract}
Large language model agents increasingly rely on reusable skills, making skill retrieval a critical front-end component of agent systems. Skill retrieval, however, is not ordinary document retrieval: a useful top-$K$ result must contain individually relevant skills that also form an executable set for the current query. Existing benchmarks and training pipelines largely supervise pairwise relevance and discard the rejection decisions produced when a language model judges a sampled skill combination to be implausible.
We introduce R3-Skill, a Chinese--English benchmark that retains these rejections as query-conditioned compatibility supervision. R3-Skill contains 10,246 deduplicated skills, 41,592 accepted queries, and 32,828 rejected annotations across four language directions; all multi-skill test labels were independently reviewed by nine experts, and 15,962 parseable rejections are organized into an eight-class taxonomy.
We further propose a two-stage system composed of R3-Embedding, a multi-positive bi-encoder for large-pool recall, and R3-Reranker, a cross-encoder trained with graded ListNet supervision. Our analysis shows that this signal is stage-dependent, helping cross-encoder reranking while providing no benefit for the tested bi-encoder objective.
On R3-Skill, the complete pipeline achieves $75.39\%$ Hit@1, $81.97\%$ NDCG@10, and $33.27\%$ Set-Compat, a $36.6\%$ relative gain over the strongest reranking baseline. It also obtains $83.87\%$ NDCG@10 on SkillRet, demonstrating transfer beyond R3-Skill.
\end{abstract}

\section{Introduction}

Reusable skills package instructions, constraints, scripts, and tool-use procedures into capabilities that an agent can retrieve and execute. This abstraction extends earlier tool-use systems such as ReAct~\cite{ref1} and Toolformer~\cite{ref2}: whereas a tool exposes an atomic call, a skill describes how to solve a class of tasks, often through a multi-step procedure. The Agent Skills specification~\cite{ref3} and recent skill ecosystems~\cite{ref4,ref5,ref6} have made this abstraction increasingly common. Skills can also function as updatable agent memory; for example, Memento-Skills reports substantial relative gains on GAIA and HLE~\cite{ref7}.

\begin{figure}[t]
\begin{minipage}[t]{0.48\columnwidth}
\centering
\includegraphics[width=\linewidth]{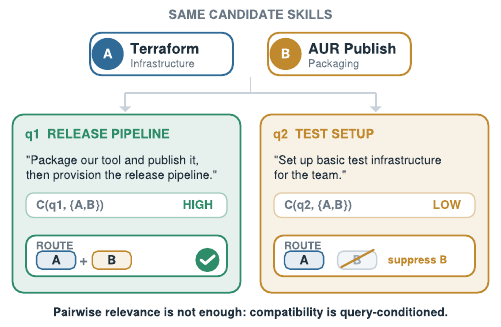}
\caption{Query-conditioned skill compatibility. The same candidate skills are jointly necessary for $q_1$ but redundant for $q_2$. Pairwise relevance alone cannot capture this query-conditioned set decision.}
\label{fig:compatibility}
\end{minipage}\hfill
\begin{minipage}[t]{0.48\columnwidth}
\centering
\includegraphics[width=\linewidth]{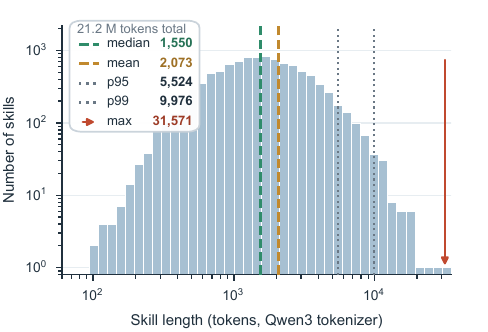}
\caption{Skill-length distribution in R3-Skill, measured over names, descriptions, and bodies with the Qwen3 tokenizer. The distribution has a long tail: the median is 1,550 tokens and the maximum is 31,571.}
\label{fig:skill_token_dist}
\end{minipage}
\end{figure}

Large skill libraries make an explicit routing layer necessary. Loading every skill into the prompt is expensive and eventually impossible: public collections already contain tens or hundreds of thousands of entries~\cite{ref8,ref9,ref10}, while a single skill may include a long body of API constraints, dependencies, examples, and canonical code. In R3-Skill, the complete library contains 21.2M tokens, and the longest skill contains 31,571 tokens (Figure~\ref{fig:skill_token_dist}). Moreover, metadata alone is insufficient when many skills have similar names or descriptions; SkillRouter reports absolute drops of $31$--$44$ points when the body is removed from routing inputs~\cite{ref13}. A practical agent must therefore retrieve a small set of skills from full, heterogeneous skill documents.

The central difficulty is that retrieved skills are acted upon rather than merely read. In document retrieval, returning two relevant passages together is usually harmless. In skill routing, two individually plausible skills may be redundant, contradictory, or impossible to compose for the current request. Figure~\ref{fig:compatibility} illustrates this query dependence: the same pair may be jointly necessary for one task but unnecessary for another. Multi-skill benchmarks also show that adding skills does not uniformly improve task completion~\cite{ref14,ref15,ref16}. Thus, successful routing requires both \emph{relevance} between each query--skill pair and \emph{compatibility} of the retrieved set under the query.

Existing skill-retrieval resources emphasize relevance supervision. End-to-end benchmarks provide limited retriever training data, whereas retrieval benchmarks such as SkillRet~\cite{ref11} and SkillRouter~\cite{ref13} inherit the standard query--candidate formulation. Meanwhile, language-model-based data synthesis already computes a useful additional signal: before writing a query for a sampled skill set, the generator often rejects combinations that do not form a natural task. These rejection decisions are normally discarded. We instead treat them as supervision about what should \emph{not} be retrieved together under a query. Concurrent work on Capability Pages~\cite{ref28} rewrites skill documents offline to expose neighbor-relative boundaries; we instead keep documents fixed and supervise set compatibility for retrieve--rerank models.
We introduce \textbf{R3-Skill} and the \textbf{Reject-as-Resource Retriever (R3)}. R3-Skill retains the WRITE/SKIP judgments, covers four Chinese--English directions, separates train and test skills, and provides expert-verified test labels. R3 uses a bi-encoder for large-pool recall and a cross-encoder for compatibility-aware reranking. The design deliberately assigns rejection supervision to the stage that can model query--candidate interaction.

Our contributions are fourfold:
\begin{itemize}
    \item We formulate skill routing as relevance plus query-conditioned set compatibility, clarifying why independently relevant skills need not form a valid routed set.
    \item We construct R3-Skill with 10,246 skills, 41,592 accepted queries, 32,828 rejection annotations, four language directions, disjoint skill pools, and an eight-class rejection taxonomy.
    \item We develop R3-Embedding and R3-Reranker, showing that rejection supervision benefits graded listwise reranking but not the tested bi-encoder objective.
    \item The complete system reaches $75.39\%$ Hit@1, $81.97\%$ NDCG@10, and $33.27\%$ Set-Compat on R3-Skill, and transfers to the SkillRet benchmark.
\end{itemize}

\section{Related Work}

Agent research has progressed from atomic tool calls toward reusable procedural capabilities~\cite{ref1,ref2,ref5,ref6}. Recent systems organize, acquire, internalize, or evolve skill libraries~\cite{ref7,ref8,ref10,ref12,ref16}. Complementary benchmarks evaluate whether skills improve downstream task completion~\cite{ref9,ref14,ref15}. These studies establish the utility and variability of skills, but end-to-end failures conflate retrieval, composition, interpretation, and execution. R3-Skill isolates the upstream routing problem while retaining supervision about multi-skill composition.

Classical sparse and dense retrieval methods, including BM25~\cite{ref21}, DPR~\cite{ref18}, ANCE~\cite{ref19}, and RocketQA~\cite{ref20}, rank candidates primarily through query--document relevance. SkillRouter~\cite{ref13} adapts a retrieve-and-rerank pipeline to large full-text skill collections and demonstrates the importance of body content. SkillRet~\cite{ref11} provides large disjoint training and evaluation pools and shows strong gains from domain-specific fine-tuning. Unlike prior skill-retrieval resources such as SkillRet and SkillRouter, R3-Skill retains rejected multi-skill combinations and lets us study where their compatibility signal should enter a two-stage retriever. A complementary line of work, Capability Pages~\cite{ref28}, leaves the online scorer unchanged and improves routing by rewriting the offline skill library; our focus is the opposite axis---training signals and models under fixed skill documents.

Multilingual embedding models improve cross-language semantic matching~\cite{ref23}, but multilinguality alone does not provide supervision about whether several capabilities should be used together. This distinction is important because public skill ecosystems are strongly language-imbalanced~\cite{ref22}. Related work also identifies structural alignment as a source of retrieval-augmented agent failures~\cite{ref17}. R3-Skill combines bilingual routing with explicit WRITE/SKIP judgments, enabling separate analysis of semantic relevance and query-conditioned compatibility.

\section{Problem Formulation}

Let $\mathcal{S}$ be a skill library and let $S_q^*\subseteq\mathcal{S}$ denote the set of skills required by query $q$. A router returns an ordered list $\pi_q$ and succeeds only when the required skills appear near the top. Pairwise relevance is necessary, but it does not fully describe a multi-skill result: the selected skills must also form a coherent task-specific set. We use the following decomposition as a modeling view:
{\small
\[
\Pr[\,S_q^* \subseteq \text{top-}K \mid q\,] \approx
\underbrace{\prod_{s \in S_q^*} \Pr[\,s \in \text{top-}K \mid q\,]}_{\text{pairwise relevance}}
\cdot
\underbrace{C(q,S_q^*)}_{\text{set compatibility}},
\]}
where $C(q,S_q^*)\in\mathbb{R}_{\geq 0}$ is a query-conditioned correction; $C\equiv1$ recovers an independent-relevance view. R3-Skill operationalizes this missing factor through two generator outcomes. WRITE indicates that a sampled skill set supports a natural query, whereas SKIP records that the set should not be jointly selected for that query context. This formulation motivates two requirements. First, evaluation should measure complete set recovery, not only per-item recall. Second, the model component consuming SKIP supervision must represent query-dependent interactions. A bi-encoder assigns each skill a shared vector across all queries, while a cross-encoder can condition each candidate score on the full query. We test this architectural distinction directly.

\section{R3-Skill Benchmark}

\paragraph{Task and design goals.} R3-Skill isolates skill routing as a standalone retrieval problem. Each evaluation item consists of a user query $q$, a held-out skill library, and a ground-truth set $S_q^*$ containing one to three skills. The router must rank the required capabilities above thousands of alternatives without observing downstream execution. This separation is deliberate: end-to-end task failure can arise from retrieval, composition, interpretation, or execution, whereas R3-Skill measures the upstream decision directly. The benchmark is built around four goals. First, it tests \emph{full-text routing}: candidates contain a name, description, and long procedural body rather than a short tool signature. Second, it tests \emph{multi-skill sufficiency}: 15,791 accepted queries require two or three skills, so success depends on recovering a coherent set. Third, it tests \emph{cross-lingual routing} across en2en, en2zh, zh2en, and zh2zh. Finally, it preserves \emph{hard rejected combinations}. These SKIPs are sampled from nearby skills and therefore distinguish query-conditioned incompatibility from trivial topical mismatch; Table~\ref{tab:r3_examples} shows representative WRITE and SKIP cases.

\paragraph{Scope and positioning.} R3-Skill targets large registries in which many candidates overlap semantically and a request may require a small skill set. It is not intended to estimate the natural prevalence of incompatible combinations in live traffic: the sampling procedure deliberately concentrates on difficult neighborhoods. Relative to SkillRouter~\cite{ref13} and SkillRet~\cite{ref11}, R3-Skill has a smaller skill pool than the former and fewer training queries than the latter, but adds four bilingual directions, 32,828 retained rejection annotations, and an eight-class rejection taxonomy. These resources make it possible to study not only whether a retriever finds relevant skills, but also where compatibility supervision should enter the model.

\begin{table}[t]
\centering
\small
\setlength{\tabcolsep}{1.5pt}
\begin{tabular}{@{}p{0.36\columnwidth}cp{0.40\columnwidth}@{}}
\toprule
Skill combination & Decision & \multicolumn{1}{c}{Why} \\
\midrule
Terraform IaC + AUR Publish & WRITE & Natural release and infrastructure pipeline \\
Code Review + Systematic Code Review & SKIP & Redundant capabilities; one skill is sufficient \\
Rust Reliability + Go Security Patterns & SKIP & Inconsistent language ecosystems for one request \\
\bottomrule
\end{tabular}
\caption{Illustrative R3-Skill decisions. SKIP sets remain topically plausible but are rejected because their skills are redundant or cannot form one natural request.}
\label{tab:r3_examples}
\end{table}

These decisions are query-conditioned: the same capability can be jointly necessary in one request yet redundant or unsuitable in another.

\subsection{Skill Collection and Deduplication}

We collect 95,212 raw skills from public GitHub repositories: 94,966 English entries and 246 Chinese entries. The raw collection contains extensive forks, mirrors, and near-identical file-operation skills. We apply cross-source filename and content hashing, remove pure file-operation entries, and retain 10,246 skills. The resulting documents preserve each skill's name, description, and body. Table~\ref{tab:skill_pool} summarizes the filtering and pool split.

\begin{table}[t]
\centering

\small
\begin{tabular}{lr}
\toprule
Stage & \#skill \\
\midrule
Raw collection (English, GitHub)         & 94,966 \\
Raw collection (Chinese, GitHub)         &    246 \\
\midrule
Total raw                                & 95,212 \\
After cross-source dedup + file-op removal & 10,246 \\
\midrule
\quad train / test pool split (80 / 20)  & 8,196 / 2,050 \\
\bottomrule
\end{tabular}
\caption{Skill pool construction. Cross-source forks and mirrors are removed by two-layer (filename + content) hash deduplication; pure file-operation entries are filtered per source.}
\label{tab:skill_pool}
\end{table}

\subsection{Compatibility-Oriented Sampling}

Uniformly sampled combinations produce mostly trivial rejections because unrelated skills are easy to identify. We instead sample semantically close combinations. BGE-M3 encodes the skill documents, after which we compare KMeans configurations with $K\in\{10,20,30,40\}$ using clustering quality and human inspection. We select 40 sub-clusters and merge them into eight thematic super-domains. Single-skill cases are sampled freely; two-skill cases draw both skills from the same sub-cluster; and three-skill cases draw all skills from the same super-domain. Neighborhoods here serve only hard WRITE/SKIP sampling for query synthesis, not offline rewriting of skill documents~\cite{ref28}.
These constraints produce nontrivial decisions. The generator accepts $56.3\%$ of two-skill combinations and $36.5\%$ of three-skill combinations, even though all sampled skills are semantically nearby under the hierarchy. Thus, semantic proximity does not imply that a combination forms a natural task.

\subsection{WRITE/SKIP Generation}

\begin{figure*}[t]
\centering
\includegraphics[width=0.98\textwidth]{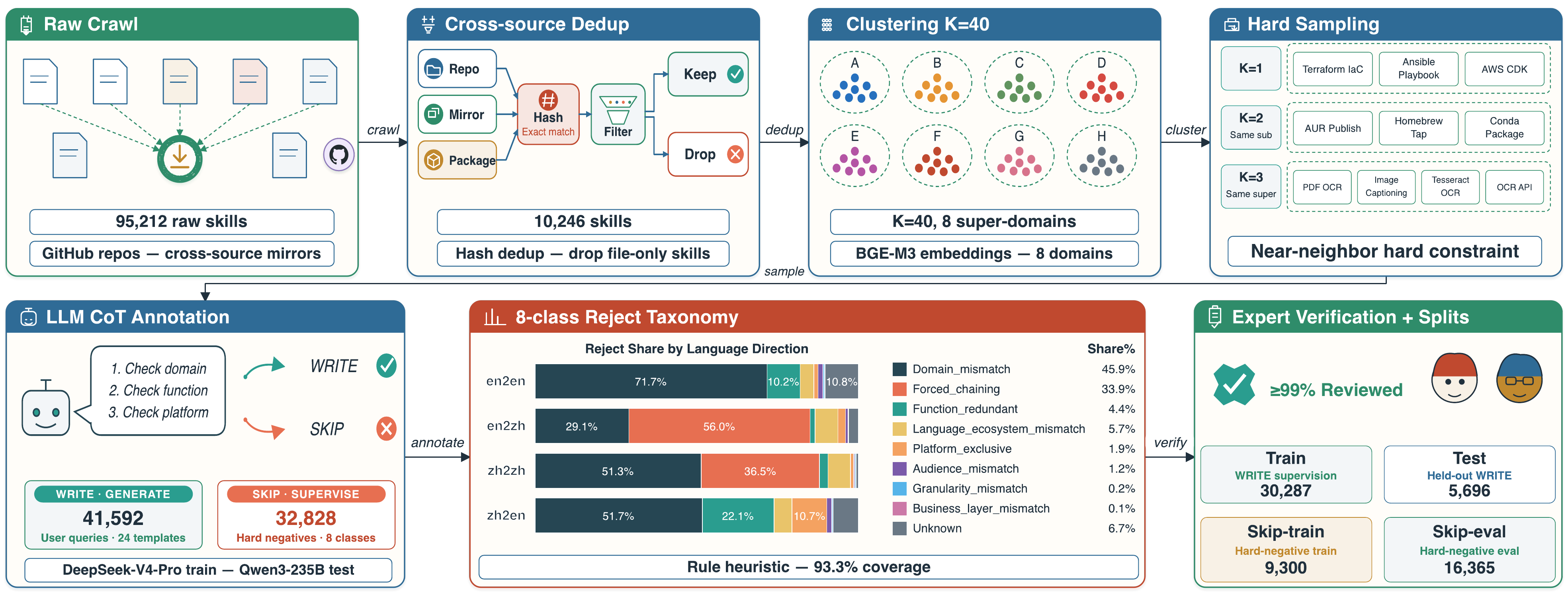}
\caption{Detailed view of the R3-Skill data production pipeline: raw 95{,}212 $\to$ cross-source dedup 10{,}246 $\to$ clustering $\to$ near-neighbor hard-constraint sampling $\to$ LLM CoT annotation (WRITE / SKIP) $\to$ multi-expert verification $\to$ benchmark splits and the 8-class Reject Taxonomy.}
\label{fig:pipeline}
\end{figure*}

Figure~\ref{fig:pipeline} presents the complete construction process. Each sampled set is first judged jointly. A WRITE decision produces a user query; a SKIP decision preserves the rejected set and its rationale. For $k=1$, invalid generations are retained only for accounting purposes. Compatibility supervision uses the $k\geq2$ rejections, for which the model explicitly judges joint plausibility before query generation.
The WRITE branch uses six request styles: \emph{task\_direct}, \emph{scene}, \emph{role\_setup}, \emph{constraint}, \emph{multi\_step}, and \emph{debug}. Generation covers four directions: en2en, en2zh, zh2en, and zh2zh. Table~\ref{tab:query_distribution} reports the accepted-query distribution by direction and number of required skills.

DeepSeek-V4-Pro~\cite{ref24} generates training queries and SKIP annotations. Qwen3-235B-A22B~\cite{ref25} drafts the test queries and labels. We deliberately use different model families for training and evaluation to reduce the risk that a retriever benefits from a single generator's recurring phrasing. The full prompts specify the requested language, one of six user styles, a 50--300 word or character range, and explicit prohibitions against skill-name disclosure and sentence copying.

\begin{table}[!h]
\centering

\small
\begin{tabular}{lrrrr}
\toprule
Direction & \#query & k=1 & k=2 & k=3 \\
\midrule
en2en & 21,332 & 12,227 & 6,535 & 2,570 \\
en2zh & 17,957 & 12,326 & 4,674 & 957 \\
zh2en & 1,148 & 641 & 353 & 154 \\
zh2zh & 1,155 & 607 & 378 & 170 \\
Total & 41,592 & 25,801 & 11,940 & 3,851 \\
\bottomrule
\end{tabular}
\caption{Query distribution by direction \& by k.}
\label{tab:query_distribution}
\end{table}

\subsection{Rejection Taxonomy and Quality Control}

We derive an eight-class taxonomy from 15,962 SKIP records with parseable rationales. The rules match specific causes first and use \emph{forced\_chaining} as a final fallback for rationales describing an artificial combination of otherwise plausible skills. Rule coverage is $93.3\%$; the remaining $6.7\%$ is labeled \emph{unknown}. Taxonomy proportions are computed only on this parseable subset and exclude single-skill generation failures.

The taxonomy exposes language-dependent rejection patterns. en2en rejections are dominated by domain mismatch, whereas $56.0\%$ of en2zh rejections are assigned to forced chaining. For en2zh with $k=3$, the SKIP rate reaches $80.9\%$. Under the strictest three-skill sampling condition, $63.5\%$ of combinations are still rejected, showing that nearest-neighbor sampling remains far from sufficient for compatibility.

\subsection{Splits and Benchmark Scope}

We split at the skill level: 8,196 skills form the training pool and 2,050 form the test pool, with no overlap. The final training split contains 30,287 accepted queries and the test split contains 5,696. We discard 5,609 queries whose required skills cross the two pools, preventing a model from memorizing test capabilities during training.

For quality assurance, all multi-skill test labels are independently reviewed by nine experts. The initial disagreement rate was below $5\%$; disputed cases underwent expert re-review followed by a second round of independent scoring. In addition, $25\%$ of the test queries are sampled for naturalness review. Automated checks find no duplicate final queries. A conservative skill-name substring audit yields leakage rates of $7.8\%$ for en2en, $2.3\%$ for en2zh, $2.1\%$ for zh2zh, and $0.0\%$ for zh2en; these matches include legitimate technology terms and are reported as an audit rather than treated as proof of semantic leakage. We also spot-check taxonomy assignments and obtain at least $99\%$ correctness on the inspected rule hits. Appendix~\ref{app:dataset_stats} reports the complete checklist and taxonomy distributions, and Appendix~\ref{app:gen-prompts} lists the prompt templates.

Together, the disjoint pools, cross-generator construction, expert review, and explicit leakage audit mitigate several risks: memorization, generator-style overfitting, label error, and lexical shortcuts. They do not make synthetic requests equivalent to live traffic, but they provide a controlled benchmark for studying large-pool, multilingual, and compatibility-aware routing.

\section{Method}

R3 follows a retrieve-and-rerank architecture. \textbf{R3-Embedding}, initialized from Qwen3-Embedding-0.6B~\cite{ref27}, retrieves 20 candidates from the full skill pool. \textbf{R3-Reranker}, initialized from Qwen3-Reranker-0.6B~\cite{ref27}, then reranks these candidates. The stages divide the problem by capacity: the bi-encoder provides efficient global recall, while the cross-encoder resolves fine-grained, query-dependent competition.

\subsection{R3-Embedding: Multi-Positive Recall}

For query $q$ with ground-truth set $S_q^*$, we sample one main positive $s^+\in S_q^*$ and mine one offline hard negative $s^-$ using Qwen3-Embedding-8B (rank in $[20,50)$ and cosine similarity below 0.85). Positives from other queries in the batch act as additional negatives. Let
$z_c=\mathbf{e}(q)^\top\mathbf{e}(c)/\tau$,
$\mathcal{C}_q=\{s^+,s^-\}\cup\text{in-batch}$, and
$S_q^\diamond=S_q^*\setminus\{s^+\}$. To avoid treating sibling ground-truth skills as negatives, we define
$\widetilde{\mathcal{C}}_q=\mathcal{C}_q\setminus S_q^\diamond$ and
$\mathcal{N}_q=\mathcal{C}_q\setminus(\{s^+\}\cup S_q^\diamond)$.
The objective is
\[
\mathcal{L}_{\mathrm{emb}}(q)=\mathcal{L}_{\mathrm{nce}}(q)+\lambda\mathcal{L}_{\mathrm{sib}}(q),
\]
\[
\mathcal{L}_{\mathrm{nce}}(q)=-\log\frac{\exp(z_{s^+})}{\sum_{c\in\widetilde{\mathcal{C}}_q}\exp(z_c)},
\]
\[
\mathcal{L}_{\mathrm{sib}}(q)=-\frac{1}{|S_q^\diamond|}\sum_{s^\diamond\in S_q^\diamond}
\log\frac{\exp(z_{s^\diamond})}{\exp(z_{s^\diamond})+\sum_{c\in\mathcal{N}_q}\exp(z_c)}.
\]
We use $\tau=1/30$ and $\lambda=0.25$. For single-skill queries, $S_q^\diamond=\emptyset$ and the objective reduces to standard InfoNCE~\cite{ref26}. For multi-skill queries, the sibling term pulls the query toward every required skill without treating sibling positives as in-batch negatives.

\begin{table*}[t]
\centering
\small
\setlength{\tabcolsep}{2.0pt}
\begin{tabular}{@{}lrrrrrrrrrrc@{}}
\toprule
Model & H@1 & N@5 & N@10 & N@15 & R@5 & R@10 & R@15 & C@5 & C@10 & C@15 & SC \\
\midrule
\multicolumn{12}{@{}l}{\textit{Embedding}} \\
\midrule
BM25 & 6.04 & 8.00 & 8.82 & 9.25 & 9.93 & 12.41 & 14.01 & 9.32 & 11.69 & 13.18 & 0.40 \\
BGE-M3 (base) & 39.59 & 46.94 & 49.20 & 50.27 & 53.88 & 60.72 & 64.67 & 51.62 & 58.55 & 62.57 & 6.93 \\
Qwen3-Embedding-0.6B (base) & 43.12 & 51.58 & 53.72 & 54.74 & 59.44 & 65.82 & 69.56 & 57.18 & 63.54 & 67.40 & 4.16 \\
Qwen3-Embedding-8B (base) & 62.29 & 71.05 & 73.01 & 73.74 & 79.13 & 84.98 & 87.65 & 76.56 & 82.79 & 85.71 & 10.69 \\
SkillRouter-Embedding-0.6B & 22.09 & 29.51 & 32.14 & 33.54 & 36.45 & 44.48 & 49.67 & 34.94 & 42.78 & 47.82 & 2.18 \\
SkillRet-Embedding-0.6B & 46.72 & 55.86 & 58.04 & 59.10 & 64.47 & 71.01 & 74.92 & 62.31 & 69.10 & 73.03 & 11.88 \\
R3-Embedding & 72.07 & 80.06 & 81.60 & 82.05 & \textbf{87.22} & 91.77 & 93.38 & \textbf{85.31} & 90.40 & 92.22 & 28.12 \\
\midrule
\multicolumn{12}{@{}l}{\textit{+ Rerank (upstream = R3-Embedding, top-20)}} \\
\midrule
+ Qwen3-Reranker-0.6B & 68.08 & 77.61 & 79.54 & 80.11 & 86.45 & \textbf{92.12} & \textbf{94.12} & 84.46 & \textbf{90.70} & \textbf{93.05} & 17.82 \\
+ SkillRouter-Reranker-0.6B & 67.98 & 75.57 & 77.65 & 78.63 & 82.57 & 88.77 & 92.34 & 80.62 & 87.20 & 91.03 & 24.36 \\
+ R3-Reranker & \textbf{75.39} & \textbf{80.40} & \textbf{81.97} & \textbf{82.74} & 85.23 & 89.94 & 92.76 & 83.41 & 88.48 & 91.52 & \textbf{33.27} \\
\bottomrule
\end{tabular}
\caption{Complete results on R3-Skill (\%). H@1 is Hit@1; N@$k$, R@$k$, C@$k$, and SC denote NDCG@$k$, Recall@$k$, Comp@$k$, and Set-Compat. The reranking block uses the same R3-Embedding top-20 candidates. Bold marks the best result in each column among the displayed systems.}
\label{tab:r3_main}
\end{table*}

\subsection{R3-Reranker: Rejection-Aware Listwise Ranking}

For each WRITE query, SKIP partners are drawn from the same clustered neighborhood as the accepted skill set, making them topically close alternatives to the required capabilities. This construction tests whether the reranker can distinguish joint necessity from topical proximity. The reranker training list contains the ground-truth skills, these SKIP partners, and top-ranked candidates from R3-Embedding, padded to 10 candidates. We assign graded labels and optimize ListNet:
{\small
\[
\operatorname{label}(s)=
\begin{cases}
3, & s\in S_q^*,\\
1, & s\text{ is a SKIP partner},\\
0, & \text{otherwise},
\end{cases}
\]
\[
\mathcal{L}_{\mathrm{rrk}}(q)=-\sum_s y_s\log p_s.
\]}
where $y_s$ is the softmax-normalized target label and $p_s$ is the softmax of the reranker scores. SKIP partners are never treated as evaluation ground truth; the intermediate label only shapes their relative position within the training list.
This placement is intentional. A bi-encoder must map each skill to one query-independent vector, so a rejection that is valid for one query can conflict with positive evidence from another. A cross-encoder instead scores each $(q,s)$ pair jointly and can use a rejected candidate as query-specific evidence for ranking nearby capabilities below the required skills.

\section{Experiments}

\subsection{Evaluation Setup}

We ask whether task-specific training improves large-pool recall, whether rejection-aware reranking improves complete-set recovery, and whether the learned models transfer across language directions and benchmarks. We evaluate on R3-Skill and SkillRet~\cite{ref11}. Baselines include BM25~\cite{ref21}, BGE-M3~\cite{ref23}, Qwen3 models~\cite{ref25,ref27}, and released SkillRouter~\cite{ref13} and SkillRet checkpoints.

 All systems receive the same retrieval instruction. Queries are truncated to 512 tokens and skill documents to 4,096 tokens. R3-Embedding retrieves 20 candidates, matching the candidate budget used by SkillRouter and SkillRet, after which each reranker reorders the same pool. We report Hit@1, NDCG@$K$, Recall@$K$, and Comp@$K$ under the same definitions as SkillRet. We further introduce \textbf{Set-Compat}, computed only when $|S_q^*|\geq2$:
{\small
\[
\operatorname{SC}(q)=\mathbb{I}\!\left[S_q^*\subseteq\operatorname{top}_{|S_q^*|}(\pi_q)\right].
\]}
Set-Compat requires all annotated jointly necessary skills within the first $|S_q^*|$ ranks and is therefore stricter than recall at a larger cutoff. It rewards ranking the complete required set ahead of optional nearby skills, which is the practical requirement for forming an action plan from a short list.

\subsection{Main Results on R3-Skill}

Table~\ref{tab:r3_main} separates large-pool retrieval from top-20 reranking. At the embedding stage, R3-Embedding achieves $72.07\%$ Hit@1 and $81.60\%$ NDCG@10, improving over the strongest base encoder, Qwen3-Embedding-8B, by $9.78$ and $8.59$ points, respectively. It also raises Recall@10 from $84.98\%$ to $91.77\%$. Thus, routing-specific supervision and sibling-positive handling let the 0.6B model improve both early precision and candidate coverage beyond the 8B base.


External skill retrievers transfer unevenly to R3-Skill: SkillRet-Embedding beats the general-purpose 0.6B Qwen3 but trails the 8B base on Hit@1, and SkillRouter-Embedding is weak at all cutoffs. This reflects a benchmark mismatch---R3-Skill's bilingual queries, held-out skill pools, and multi-skill compatibility decisions are absent from their training resources, whereas R3-Embedding is trained on the benchmark's full-text, multi-positive structure.

Generic rerankers preserve some large-cutoff recall but reduce Hit@1 and Set-Compat. R3-Reranker instead raises Hit@1 to $75.39\%$ and Set-Compat to $33.27\%$, relative gains of $4.6\%$ and $18.3\%$ over R3-Embedding, while keeping NDCG@10 nearly unchanged. The absolute Set-Compat value is modest by design: complete-set recovery is a deliberately hard target, and no baseline exceeds $25\%$. The generic reranker retains slightly higher Recall@10--15, but Set-Compat---not large-cutoff recall---is the deployment-relevant quantity: an agent loads only a short prefix, so ranking the complete required set into the top $|S_q^*|$ positions matters more than coverage at rank 15.

\subsection{Cross-Lingual Compatibility Analysis}

The rejection taxonomy suggests two regimes. en2en SKIPs usually reflect domain-level mismatches, which a bi-encoder can represent as semantic distance. en2zh SKIPs more often describe forced chaining: the skills may be plausible in isolation or in other contexts, but do not form a jointly necessary set for the current query.

\begin{table}[b]
\centering
\small
\setlength{\tabcolsep}{2.0pt}
\begin{tabular}{lrrrrc}
\toprule
Direction & \multicolumn{2}{c}{Hit@1} & \multicolumn{2}{c}{Set-Compat} & $\Delta$SC \\
\cmidrule(lr){2-3}\cmidrule(lr){4-5}
& Emb. & +Rerank & Emb. & +Rerank & \\
\midrule
en2en & 73.41 & 75.07 & 32.87 & 36.33 & +3.46 \\
en2zh & 69.91 & 74.95 & 18.97 & 25.64 & +6.67 \\
zh2en & 86.67 & 93.33 & 45.45 & 63.64 & +18.19 \\
zh2zh & 79.35 & 77.17 & 50.00 & 60.00 & +10.00 \\
\bottomrule
\end{tabular}
\caption{Cross-lingual breakdown on R3-Skill (\%). ``Emb.'' is R3-Embedding and ``+Rerank'' adds R3-Reranker.}
\label{tab:crosslingual}
\end{table}

Table~\ref{tab:crosslingual} supports this compatibility interpretation for the two high-volume directions. At the embedding stage, en2zh Set-Compat is $18.97\%$, below en2en at $32.87\%$. R3-Reranker improves Set-Compat by $6.67$ points on en2zh, compared with $3.46$ points on en2en, consistent with the prevalence of forced chaining in en2zh. The smaller zh2en and zh2zh slices are less stable, so we treat them as diagnostic rather than confirmatory evidence.

\subsection{Cross-Benchmark Transfer}

\begin{table*}[t]
\centering
\small
\setlength{\tabcolsep}{2.0pt}
\begin{tabular}{@{}lrrrrrrrrr@{}}
\toprule
Model & N@5 & N@10 & N@15 & R@5 & R@10 & R@15 & C@5 & C@10 & C@15 \\
\midrule
\multicolumn{10}{@{}l}{\textit{Embedding}} \\
\midrule
BM25 & 46.47 & 48.86 & 49.90 & 50.09 & 56.55 & 59.95 & 34.96 & 41.09 & 44.27 \\
Qwen3-Embedding-0.6B (base) & 56.23 & 58.35 & 59.34 & 59.29 & 64.89 & 68.06 & 41.98 & 47.27 & 50.31 \\
Qwen3-Embedding-8B (base) & 57.57 & 59.98 & 61.05 & 60.71 & 67.06 & 70.45 & 43.33 & 50.01 & 53.69 \\
SkillRouter-Embedding-0.6B & 68.39 & 70.38 & 71.22 & 70.50 & 75.63 & 78.28 & 52.79 & 59.04 & 62.26 \\
SkillRet-Embedding-0.6B & 75.57 & 78.03 & 78.87 & 79.15 & 85.42 & 88.09 & 65.96 & 75.09 & 79.03 \\
R3-Embedding & 78.79 & 81.06 & 81.89 & 81.82 & 87.64 & 90.20 & 68.78 & 77.43 & 81.33 \\
\midrule
\multicolumn{10}{@{}l}{\textit{+ Rerank (upstream = R3-Embedding, top-20)}} \\
\midrule
+ Qwen3-Reranker-0.6B & 73.24 & 76.03 & 76.97 & 78.34 & 85.69 & 88.70 & 64.52 & 74.28 & 78.47 \\
+ SkillRouter-Reranker-0.6B & 80.49 & 82.15 & 82.71 & 82.98 & 87.30 & 89.11 & 70.46 & 76.33 & 78.89 \\
+ R3-Reranker & \textbf{82.41} & \textbf{83.87} & \textbf{84.38} & \textbf{85.47} & \textbf{89.36} & \textbf{91.06} & \textbf{75.48} & \textbf{81.01} & \textbf{83.15} \\
\bottomrule
\end{tabular}
\caption{Complete transfer results on SkillRet (\%). Horizontal rules separate embedding retrieval from top-20 reranking with R3-Embedding as the shared upstream model.}
\label{tab:skillret_transfer}
\end{table*}

\begin{table}[b]
\centering
\small
\setlength{\tabcolsep}{5.0pt}
\begin{tabular}{llrrrc}
\toprule
Stage & Negative source & H@1 & N@10 & R@10 & SC \\
\midrule
\multicolumn{6}{l}{\textit{Embedding (InfoNCE)}} \\
A & 8B-mined & \textbf{72.07} & \textbf{81.60} & \textbf{91.77} & \textbf{28.12} \\
B & SKIP if available & 69.93 & 79.94 & 90.90 & 21.58 \\
\midrule
\multicolumn{6}{l}{\textit{Reranker (listwise)}} \\
C & no SKIP labels & 72.31 & 81.09 & \textbf{91.08} & 29.50 \\
D & shuffled SKIP labels & 75.11 & 81.93 & 89.78 & 29.70 \\
E & SKIP labels & \textbf{75.39} & \textbf{81.97} & 89.94 & \textbf{33.27} \\
\bottomrule
\end{tabular}
\caption{Stage-wise SKIP ablation on R3-Skill (\%). D shuffles E's matched SKIP partners; A/E are deployed.}
\label{tab:ablation}
\end{table}

Table~\ref{tab:skillret_transfer} evaluates transfer to an independently constructed English benchmark. At the embedding stage, R3-Embedding reaches $81.06\%$ NDCG@10, exceeding SkillRet-Embedding-0.6B by $3.03$ points and SkillRouter-Embedding-0.6B by $10.68$ points. The gain is consistent across every cutoff and metric family. In particular, R3-Embedding improves Comp@10 from $75.09\%$ to $77.43\%$ over the SkillRet-trained encoder, showing that R3-Embedding transfers to complete-set retrieval even though SkillRet does not provide the same rejection supervision.

The result also separates task adaptation from scale. Qwen3-Embedding-8B reaches only $59.98\%$ NDCG@10, whereas all three skill-specific encoders perform substantially better. R3-Embedding's improvement is therefore unlikely to come from backbone capacity; it reflects the retrieval data, full skill representation, and multi-skill objective. 


Reranking exhibits a clear domain-adaptation hierarchy. The generic Qwen3 reranker lowers every NDCG metric relative to R3-Embedding and also hurts completeness, showing a general pair scorer can override a strong domain-specific retriever. SkillRouter-Reranker improves early NDCG but slightly reduces Recall@10 and Comp@10, so its ordering preference transfers only partially. R3-Reranker improves all nine metrics, reaching $83.87\%$ NDCG@10, $89.36\%$ Recall@10, and $81.01\%$ Comp@10. Unlike the R3-Skill result, no recall trade-off appears on SkillRet---the reranker improves both early ordering and set coverage.

This cross-benchmark pattern strengthens the stage-wise interpretation. On R3-Skill, SKIP supervision mainly changes the smallest sufficient prefix and yields the largest gain in Set-Compat. On SkillRet, where evaluation uses larger-cutoff ranking and completeness metrics, the same reranker also improves broad coverage. The combined evidence shows that R3-Reranker's gains transfer to an evaluation setting without rejection annotations.

\subsection{Where Should SKIP Supervision Enter?}
\label{sec:ablation}


Table~\ref{tab:ablation} gives a clear stage-wise answer. With all else fixed, B substitutes a SKIP partner when available and otherwise keeps A's 8B-mined negative; it drops Hit@1 from $72.07\%$ to $69.93\%$ and Set-Compat from $28.12\%$ to $21.58\%$. D matches E in language direction, required-set size, candidate similarity, skill length, and training exposure, but assigns label-1 candidates to different queries; relative to C it raises Hit@1/NDCG@10 from $72.31\%$/$81.09\%$ to $75.11\%$/$81.93\%$. Using the original query--SKIP pairing in E leaves these nearly unchanged but lifts Set-Compat from $29.70\%$ to $33.27\%$. The C/D and D/E contrasts thus separate the benefit of adding matched hard candidates from that of preserving correct query--SKIP alignment, supporting SKIP labels for reranking rather than bi-encoder negatives.

\paragraph{Why SKIP conflicts with the bi-encoder.} The embedding degradation is structural. For a SKIP partner $s'$ under $q$, the InfoNCE gradient is $+p_{s'}\mathbf{e}(q)/\tau$; when $s'$ is positive for another query $\tilde q$, it is $-(1-\tilde p_{s'})\mathbf{e}(\tilde q)/\tau$. One shared vector must therefore reconcile a query-specific push with positive pulls elsewhere. The cross-encoder avoids this shared-vector constraint by recomputing a representation for each $(q,s)$ pair; Appendix~\ref{app:bilateral} gives the derivation.

\section{Discussion}

\paragraph{Empirical findings and interpretation.} 
A SKIP judgment does not mean every skill in the rejected set is globally irrelevant; it indicates the set is not jointly necessary for a given request. Treating this as a conventional contrastive negative asks a shared embedding to encode mutually inconsistent relations, whereas graded cross-encoder supervision lets the score vary with the query, aligning the model with the annotation semantics. The rejection taxonomy adds a diagnostic layer beyond aggregate accuracy: domain mismatch behaves like a conventional hard negative, while forced chaining more directly captures query-conditioned incompatibility. The larger en2zh Set-Compat gain fits this distinction, though the small zh2en and zh2zh subsets remain too limited for a strong language-specific claim. The D/E control sharpens this: matched hard candidates improve early ranking, whereas preserving the original query pairing yields the larger Set-Compat gain—linking rejection annotations to the short-list decision that matters when an agent assembles a small action plan.

\paragraph{Deployment trade-offs.} 
Deployment need not always invoke both stages. On the complete 10,246-skill collection, reranking 20 candidates ($0.20\%$ of the pool) cuts query--skill comparisons by about $512\times$: at the evaluation caps, the 20 pairs require at most $20\times(512+4{,}096)=92{,}160$ input tokens per request, versus $10{,}246\times(512+4{,}096)\approx47.2$ million for full-pool scoring. These are capped input volumes, not measured latency, which also depends on hardware, batching, precision, indexing, and realized lengths. Use R3-Embedding alone for broad shortlist coverage, R3-Reranker for a prefiltered pool, and the full pipeline when both are required.

\paragraph{Why Set-Compat matters in deployment.} 
An agent rarely loads a long skill list: each added skill consumes context and may introduce redundant or incompatible instructions. Set-Compat asks whether the annotated jointly necessary skills fit within the smallest actionable prefix. A set recovered only at ranks 10--15 can lift long-cutoff coverage yet rarely enters an action plan built from the leading candidates. In Table~\ref{tab:r3_main}, R3-Embedding gives the strongest short-list coverage—highest Recall@5 ($87.22\%$) and Comp@5 ($85.31\%$)—while R3-Reranker attains the best Hit@1 ($75.39\%$), NDCG, and Set-Compat ($33.27\%$). The SC gain thus captures an ordering improvement directly useful under a small execution budget.

\section{Limitations}
Our findings should be interpreted within R3-Skill's current scope: queries are synthetic, evaluation is offline, non-English skill pools remain small, the rejection taxonomy is rule-based, and reranking is limited to the top-20 candidates. Future work should evaluate larger multilingual registries, live user requests, and end-to-end agent execution.

\section{Conclusion}


We argued that skill routing differs from document retrieval because a successful result must satisfy both per-skill relevance and query-conditioned set compatibility. R3-Skill makes this distinction measurable by retaining the language-model rejection decisions that synthesis pipelines normally discard, and R3 separates efficient multi-positive recall from rejection-aware listwise reranking. Our central finding is not that rejection supervision is harmful, but that it must be matched to the component able to represent its conditional structure: forced into a single shared embedding it does not help, whereas in a query-conditioned cross-encoder it yields the largest gains at top ranks and in complete-set recovery, and these gains transfer to SkillRet, which carries no rejection annotations. More broadly, a generator's implicit judgment about what should \emph{not} be used together is recoverable as first-class supervision. We will release code and data upon publication.

\clearpage

\bibliographystyle{arxivstyle}
\bibliography{references}

\appendix
\section{Evaluation Details}
\label{app:eval-detail}

\paragraph{Metrics.} We use the same Recall@K and Comp@K definitions as SkillRet. Set-Compat is the stricter all-ground-truth indicator $\mathbb{1}[S_q^* \subseteq \text{top-}m]$ with $m=|S_q^*|$, evaluated only on multi-skill queries. NDCG@5/10/15 and Hit@1 follow standard definitions.

\paragraph{Lengths and prompt.} Lengths are uniformly set to: query up to 512 tokens, doc up to 4096 tokens, skill body up to 4096 tokens. All baselines and our models are evaluated under the same instruction: \texttt{Instruct: Given a user request, retrieve the agent skill that solves it.} followed by the query.

\paragraph{Reranker pool and the rank trade-off.} The reranker input pool is the top-20 from R3-Embedding ($K=20$, matching the SkillRouter / SkillRet setting). Since the pool already covers the embedding top-10, every GT in the embedding top-10 is also in the reranker candidate set; aggregate R@10 / Comp@10 movement at the reranker stage is therefore not due to candidate truncation. Internally, the reranker promotes the entire GT set toward the top, but may move some individual GT items lower in the process---the rank trade-off paid for the gain in Set-Compat. Whether this surfaces as a small aggregate R@10 dip or a small lift depends on how saturated the embedding top-10 already is on a given test set.

\section{Complete Evaluation Results}
\label{app:full-results}

The main paper reports the primary metrics needed for its claims. Tables~\ref{tab:full_r3_embedding}--\ref{tab:full_skillret_reranker} provide all cutoffs used in the evaluation. Table~\ref{tab:full_crosslingual} gives the complete language-direction breakdown.

\begin{table}[htbp]
\centering
\small
\setlength{\tabcolsep}{1.0pt}
\begin{tabular}{@{}lrrrrrrrrrrC@{}}
\toprule
Model & H@1 & N@5 & N@10 & N@15 & R@5 & R@10 & R@15 & C@5 & C@10 & C@15 & SC \\
\midrule
BM25 & 6.04 & 8.00 & 8.82 & 9.25 & 9.93 & 12.41 & 14.01 & 9.32 & 11.69 & 13.18 & 0.40 \\
BGE-M3 (base) & 39.59 & 46.94 & 49.20 & 50.27 & 53.88 & 60.72 & 64.67 & 51.62 & 58.55 & 62.57 & 6.93 \\
Qwen3-Embedding-0.6B (base) & 43.12 & 51.58 & 53.72 & 54.74 & 59.44 & 65.82 & 69.56 & 57.18 & 63.54 & 67.40 & 4.16 \\
Qwen3-Embedding-8B (base) & 62.29 & 71.05 & 73.01 & 73.74 & 79.13 & 84.98 & 87.65 & 76.56 & 82.79 & 85.71 & 10.69 \\
SkillRouter-Embedding-0.6B & 22.09 & 29.51 & 32.14 & 33.54 & 36.45 & 44.48 & 49.67 & 34.94 & 42.78 & 47.82 & 2.18 \\
SkillRet-Embedding-0.6B & 46.72 & 55.86 & 58.04 & 59.10 & 64.47 & 71.01 & 74.92 & 62.31 & 69.10 & 73.03 & 11.88 \\
R3-Embedding & \textbf{72.07} & \textbf{80.06} & \textbf{81.60} & \textbf{82.05} & \textbf{87.22} & \textbf{91.77} & \textbf{93.38} & \textbf{85.31} & \textbf{90.40} & \textbf{92.22} & \textbf{28.12} \\
\bottomrule
\end{tabular}
\caption{Embedding on R3-Skill test (\%). H@1=Hit@1, N@$k$=NDCG@$k$, C@$k$=Comp@$k$, SC=Set-Compat.}
\label{tab:full_r3_embedding}
\end{table}

\begin{table}[htbp]
\centering
\small
\setlength{\tabcolsep}{2.0pt}
\begin{tabular}{@{}lrrrrrrrrr@{}}
\toprule
Model & N@5 & N@10 & N@15 & R@5 & R@10 & R@15 & C@5 & C@10 & C@15 \\
\midrule
BM25 & 46.47 & 48.86 & 49.90 & 50.09 & 56.55 & 59.95 & 34.96 & 41.09 & 44.27 \\
Qwen3-Embedding-0.6B (base) & 56.23 & 58.35 & 59.34 & 59.29 & 64.89 & 68.06 & 41.98 & 47.27 & 50.31 \\
Qwen3-Embedding-8B (base) & 57.57 & 59.98 & 61.05 & 60.71 & 67.06 & 70.45 & 43.33 & 50.01 & 53.69 \\
SkillRouter-Embedding-0.6B & 68.39 & 70.38 & 71.22 & 70.50 & 75.63 & 78.28 & 52.79 & 59.04 & 62.26 \\
SkillRet-Embedding-0.6B & 75.57 & 78.03 & 78.87 & 79.15 & 85.42 & 88.09 & 65.96 & 75.09 & 79.03 \\
R3-Embedding & \textbf{78.79} & \textbf{81.06} & \textbf{81.89} & \textbf{81.82} & \textbf{87.64} & \textbf{90.20} & \textbf{68.78} & \textbf{77.43} & \textbf{81.33} \\
\bottomrule
\end{tabular}
\caption{Embedding on SkillRet official test (\%).}
\label{tab:full_skillret_embedding}
\end{table}

\begin{table}[htbp]
\centering
\small
\setlength{\tabcolsep}{1.0pt}
\begin{tabular}{@{}lrrrrrrrrrrC@{}}
\toprule
Reranker & H@1 & N@5 & N@10 & N@15 & R@5 & R@10 & R@15 & C@5 & C@10 & C@15 & SC \\
\midrule
\multicolumn{12}{@{}l@{}}{\textit{Upstream embedding = R3-Embedding}} \\
\midrule
(no reranker, embedding only) & 72.07 & 80.06 & 81.60 & 82.05 & \textbf{87.22} & 91.77 & 93.38 & \textbf{85.31} & 90.40 & 92.22 & 28.12 \\
SkillRouter-Reranker-0.6B & 67.98 & 75.57 & 77.65 & 78.63 & 82.57 & 88.77 & 92.34 & 80.62 & 87.20 & 91.03 & 24.36 \\
Qwen3-Reranker-0.6B & 68.08 & 77.61 & 79.54 & 80.11 & 86.45 & \textbf{92.12} & \textbf{94.12} & 84.46 & \textbf{90.70} & \textbf{93.05} & 17.82 \\
R3-Reranker & \textbf{75.39} & \textbf{80.40} & \textbf{81.97} & \textbf{82.74} & 85.23 & 89.94 & 92.76 & 83.41 & 88.48 & 91.52 & \textbf{33.27} \\
\bottomrule
\end{tabular}
\caption{Rerankers on R3-Skill test (\%) over the shared R3-Embedding top-20 pool.}
\label{tab:full_r3_reranker}
\end{table}

\begin{table}[htbp]
\centering
\small
\setlength{\tabcolsep}{2.0pt}
\begin{tabular}{@{}lrrrrrrrrr@{}}
\toprule
Reranker & N@5 & N@10 & N@15 & R@5 & R@10 & R@15 & C@5 & C@10 & C@15 \\
\midrule
\multicolumn{10}{@{}l@{}}{\textit{Upstream embedding = SkillRet-Embedding-0.6B}} \\
\midrule
(no reranker, embedding only) & 75.57 & 78.03 & 78.87 & 79.15 & 85.42 & 88.09 & 65.96 & 75.09 & 79.03 \\
Qwen3-Reranker-0.6B & 72.84 & 75.81 & 76.68 & 78.23 & 85.86 & 88.57 & 64.72 & 75.48 & 79.93 \\
Qwen3-Reranker-4B & 73.24 & 76.20 & 77.12 & 78.58 & 86.18 & 89.04 & 65.52 & 76.09 & 80.47 \\
Qwen3-Reranker-8B & 73.21 & 76.09 & 76.92 & 79.12 & 86.48 & 89.07 & 66.40 & 76.53 & 80.67 \\
SkillRouter-Reranker-0.6B & 79.98 & 81.85 & 82.37 & 82.85 & 87.64 & 89.24 & 70.66 & 78.15 & 80.75 \\
SkillRet-Reranker-0.6B & 80.71 & 82.18 & 82.66 & 83.73 & 87.61 & 89.20 & 73.28 & 78.95 & 81.09 \\
\midrule
\multicolumn{10}{@{}l@{}}{\textit{Upstream embedding = R3-Embedding}} \\
\midrule
(no reranker, embedding only) & 78.79 & 81.06 & 81.89 & 81.82 & 87.64 & 90.20 & 68.78 & 77.43 & 81.33 \\
Qwen3-Reranker-0.6B & 73.24 & 76.03 & 76.97 & 78.34 & 85.69 & 88.70 & 64.52 & 74.28 & 78.47 \\
SkillRouter-Reranker-0.6B & 80.49 & 82.15 & 82.71 & 82.98 & 87.30 & 89.11 & 70.46 & 76.33 & 78.89 \\
R3-Reranker & \textbf{82.41} & \textbf{83.87} & \textbf{84.38} & \textbf{85.47} & \textbf{89.36} & \textbf{91.06} & \textbf{75.48} & \textbf{81.01} & \textbf{83.15} \\
\bottomrule
\end{tabular}
\caption{Reranker on SkillRet official test (\%). Grouped by upstream embedding.}
\label{tab:full_skillret_reranker}
\end{table}

\begin{table}[htbp]
\centering
\small
\setlength{\tabcolsep}{2.0pt}
\begin{tabular}{@{}llrrrrrrrC@{}}
\toprule
Direction & Stage & H@1 & N@10 & R@5 & R@10 & C@5 & C@10 & C@15 & SC \\
\midrule
en2en & R3-Embedding & 73.41 & 82.14 & 87.13 & 91.78 & 84.99 & 90.34 & 91.86 & 32.87 \\
en2en & + R3-Reranker & 75.07 & 81.04 & 83.65 & 88.83 & 81.48 & 87.19 & 90.74 & 36.33 \\
en2zh & R3-Embedding & 69.91 & 80.44 & 86.71 & 91.43 & 84.97 & 90.05 & 92.29 & 18.97 \\
en2zh & + R3-Reranker & 74.95 & 82.28 & 86.20 & 90.59 & 84.68 & 89.25 & 91.96 & 25.64 \\
zh2en & R3-Embedding & 86.67 & 92.87 & 97.46 & 98.73 & 96.19 & 98.10 & 98.10 & 45.45 \\
zh2en & + R3-Reranker & 93.33 & 96.37 & 97.30 & 99.68 & 96.19 & 99.05 & 99.05 & 63.64 \\
zh2zh & R3-Embedding & 79.35 & 87.14 & 93.48 & 94.02 & 92.39 & 93.48 & 94.57 & 50.00 \\
zh2zh & + R3-Reranker & 77.17 & 84.44 & 90.22 & 92.93 & 89.13 & 92.39 & 93.48 & 60.00 \\
\bottomrule
\end{tabular}
\caption{Full cross-lingual by-direction results on R3-Skill (\%).}
\label{tab:full_crosslingual}
\end{table}

\FloatBarrier
\section{Benchmark Comparison and Quality Assurance}
\label{app:benchmark-qa}

The main paper summarizes why R3-Skill differs from existing resources and reports the most important safeguards. The tables below provide an attribute-level comparison and the complete audit checklist.

\begin{table}[htbp]
\centering
\small
\setlength{\tabcolsep}{3pt}
\begin{tabular}{@{}>{\raggedright\arraybackslash}p{0.18\textwidth}>{\centering\arraybackslash}p{0.22\textwidth}>{\centering\arraybackslash}p{0.26\textwidth}>{\centering\arraybackslash}p{0.26\textwidth}@{}}
\toprule
Attribute & SkillRouter & SkillRet & R3-Skill (ours) \\
\midrule
\#skill & $\sim$80{,}000 & 17{,}810 & 10{,}246 \\
\#WRITE query & 75 eval$^{\dagger}$ & 63{,}259 train + 4{,}997 eval & 41{,}592 \\
\#SKIP & --- & --- & 32{,}828 \\
LangDir & en & en & en2en / en2zh / zh2en / zh2zh \\
Reject Signal & $\times$ discarded & $\times$ discarded & \checkmark{} retained + 8-class taxonomy \\
Train / Test split & unreleased & no overlap & no overlap \\
\bottomrule
\end{tabular}
\caption{Dataset attribute comparison with concurrent skill retrieval datasets.}
\end{table}
{\small $^{\dagger}$SkillRouter releases only 75 eval queries; its 37,979 training pairs are unreleased.}

\begin{table}[htbp]
\centering
\small
\begin{tabular}{lp{0.62\textwidth}}
\toprule
Safeguard & Procedure and resulting figure \\
\midrule
Cross-source LLMs for train/test & Train set generated by DeepSeek-V4-Pro; test set by Qwen3-235B-A22B---avoiding shared-generator bias between training and evaluation \\
Manual test-set verification & All multi-skill GT labels independently reviewed by nine experts; initial disagreement below 5\%; disputed cases underwent expert re-review followed by a second round of independent scoring; overall query naturalness: 25\% sampled review (LLM-simulated rather than real-user validated) \\
Duplicate query detection & Verified by \texttt{distinct == n}; final duplicate count is 0 \\
Cross-lingual leakage check & Substring match of skill names against query text: en2en 7.8\% / en2zh 2.3\% / zh2zh 2.1\% / zh2en 0.0\% \\
No train/test pool overlap & Split is at the skill-pool level; skills in the test set never appear in training \\
\bottomrule
\end{tabular}
\caption{Quality assurance checks (summary of procedures already described in the main paper).}
\label{tab:qa-checks}
\end{table}

\FloatBarrier
\section{Bilateral Balancing: Derivation and Alternative Loss Forms}
\label{app:bilateral}

This appendix states Theorem~\ref{thm:bilateral}, derives its two gradients, and compares the InfoNCE candidate-pool injection with three alternative loss forms.

\subsection{Setup}

Let $\mathbf{e}:\mathcal{S}\cup\mathcal{Q}\to\mathbb{S}^{d-1}$ be an L2-normalized shared encoder with temperature $\tau>0$; the logit between query $q$ and candidate $s$ is $z_s := \mathbf{e}(q)^\top\mathbf{e}(s)/\tau$. The SKIP partner $s'$ is included alongside the GT $s^+$ and random negatives $\{s^-_j\}$ in the candidate pool $\mathcal{C}=\{s^+, \{s^-_j\}, s'\}$, with InfoNCE softmax probability $p_{s'}=\exp(z_{s'})/\sum_{c\in\mathcal{C}}\exp(z_c)$. Gradients below are taken with respect to the L2-normalized $\mathbf{e}(s')$; back-propagating further to the pre-normalization representation multiplies by the L2-normalization Jacobian but preserves the direction of the similarity change.

\begin{theorem}[InfoNCE bilateral balancing on $\mathbf{e}(s')$]
\label{thm:bilateral}
When $s'$ is a SKIP partner under query $q$,
\[
\nabla_{\mathbf{e}(s')}\,\mathcal{L}_\text{NCE}^{(q)} = +\,\frac{p_{s'}}{\tau}\,\mathbf{e}(q),
\]
pushing $\mathbf{e}(s')$ away from $\mathbf{e}(q)$. If the same $s'$ also appears as a positive under another query $\tilde q$ (with softmax probability $\tilde p_{s'}$ on its candidate pool),
\[
\nabla_{\mathbf{e}(s')}\,\mathcal{L}_\text{NCE}^{(\tilde q)} = -\,\frac{1-\tilde p_{s'}}{\tau}\,\mathbf{e}(\tilde q),
\]
pulling $\mathbf{e}(s')$ toward $\mathbf{e}(\tilde q)$.
\end{theorem}

\begin{figure}[htbp]
\centering
\includegraphics[width=0.72\textwidth]{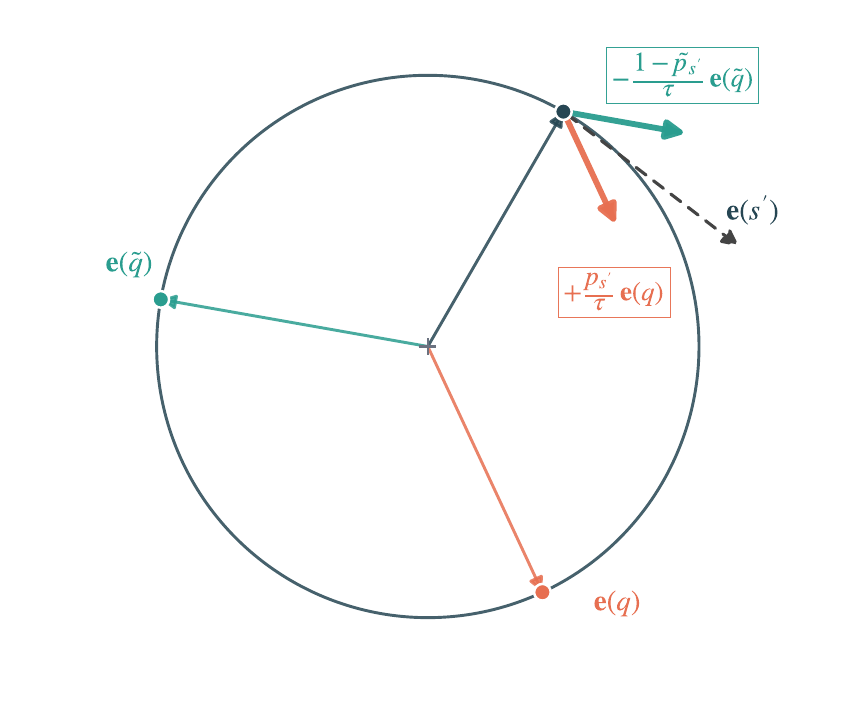}
\caption{Bilateral balancing on the shared $\mathbf{e}(s')$: a SKIP push under query $q$ is balanced by a positive pull under another query $\tilde q$; their equilibrium fixes the geometric position of $\mathbf{e}(s')$.}
\label{fig:bilateral_balance}
\end{figure}

\subsection{Derivation of Theorem~\ref{thm:bilateral}}

The InfoNCE loss on the candidate pool $\mathcal{C}$ is $\mathcal{L}_\text{NCE} = -\log p_{s^+}$. Among the logits $\{z_c\}_{c\in\mathcal{C}}$, only $z_{s'}$ depends on $\mathbf{e}(s')$, with $\partial z_{s'}/\partial \mathbf{e}(s') = \mathbf{e}(q)/\tau$. Differentiating yields
\[
\nabla_{\mathbf{e}(s')}\,\mathcal{L}_\text{NCE}^{(q)} \;=\; -\,\frac{\partial \log p_{s^+}}{\partial \mathbf{e}(s')} \;=\; +\,\frac{p_{s'}}{\tau}\,\mathbf{e}(q),
\]
the push-away gradient. When the same $s'$ appears in another sample $(\tilde q, s'^+, \ldots)$ as the positive (so $s' = s'^+$ in that sample), the relevant logit is $z_{s'}=\mathbf{e}(\tilde q)^\top\mathbf{e}(s')/\tau$, and differentiating $-\log p_{s'}$ gives
\[
\nabla_{\mathbf{e}(s')}\,\mathcal{L}_\text{NCE}^{(\tilde q)} \;=\; -\,\frac{1-\tilde p_{s'}}{\tau}\,\mathbf{e}(\tilde q),
\]
the pull-toward gradient.

The two gradients act on the same shared $\mathbf{e}(s')$ across batches: a SKIP push is weighted by $p_{s'}$, while a positive pull from $\tilde q$ is weighted by $1-\tilde p_{s'}$, which is large when $s'$ is not yet confidently ranked as the positive and vanishes as $\tilde p_{s'}\to1$. The position of $\mathbf{e}(s')$ is the equilibrium of these query-specific forces (Figure~\ref{fig:bilateral_balance}); no single query can unilaterally dominate.

\subsection{Negative Results: Limitations of Other Forms on a Bi-Encoder}

\begin{itemize}
\item Independent listwise hinge $\max(0, m - \mathbf{e}(s)\!\cdot\!\mathbf{e}(q) + \mathbf{e}(s')\!\cdot\!\mathbf{e}(q))$: computed only on the SKIP row, it provides no opposite-direction gradient on rows where $s'$ is a positive, leading to one-sided pushing.
\item BCE compatibility loss $-\log \sigma(-\mathbf{e}(s)\!\cdot\!\mathbf{e}(s'))$: the gradient direction is along $-\mathbf{e}(s)$, independent of the query, and pushes globally; it disrupts the local neighborhood structure when $s$ and $s'$ co-occur as positives under other queries.
\item Distillation form $\|\mathbf{e}(s) + \mathbf{e}(s') - \mathbf{e}_{\text{tgt}}\|^2$ (where $\mathbf{e}_{\text{tgt}}$ is a target vector from some external teacher model): the gradient direction is along $\mathbf{e}_{\text{tgt}}$, independent of the current query $\mathbf{e}(q)$, essentially pulling $\mathbf{e}(s)$ toward a fixed reference point---it lacks the ``push away $s'$ under $q$'' semantics required by SKIP.
\end{itemize}

Among the bi-encoder variants we examined, the InfoNCE candidate-pool injection is the only form that is both query-aware in its gradient direction and bilaterally balanced. However, Theorem~\ref{thm:bilateral} also bounds what it can achieve: the shared $\mathbf{e}(s')$ has to compromise between ``pushed away under $q$'' and ``pulled in under $\tilde q$'', so the SKIP push from any single query is diluted and the gain is capped by the encoder's capacity. This is why we apply the SKIP signal at the R3-Reranker (cross-encoder) stage instead.

\FloatBarrier
\section{R3-Skill Diagnostics and Dataset Statistics}
\label{app:dataset_stats}

This section expands the benchmark analysis in the main paper. It reports why $K=40$ was selected, how the 40 clusters map to eight super-domains, how WRITE/SKIP annotation proceeds, and how query style and rejection reasons vary across slices.

\begin{figure}[htbp]
\centering
\includegraphics[width=0.9\columnwidth]{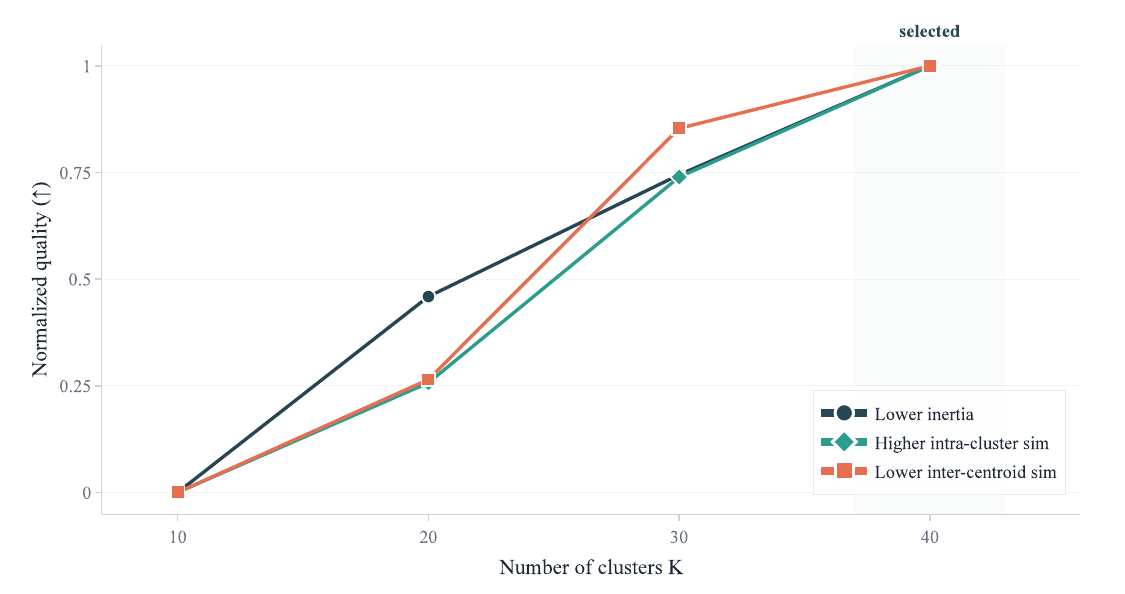}
\caption{Cluster quality vs.\ $K$ ($K \in \{10, 20, 30, 40\}$). The three indicators are inertia (within-cluster sum of squared distances, lower is better), intra-cluster sim (within-cluster topical similarity, higher is better), and inter-centroid sim (similarity between adjacent centroids, lower is better); for joint comparison, all three curves are normalized to [0,1] and re-oriented so that ``higher is better''. $K=40$ is selected based on the joint criteria.}
\label{fig:cluster_quality}
\end{figure}

\begin{figure*}[htbp]
\centering
\includegraphics[width=0.7\textwidth]{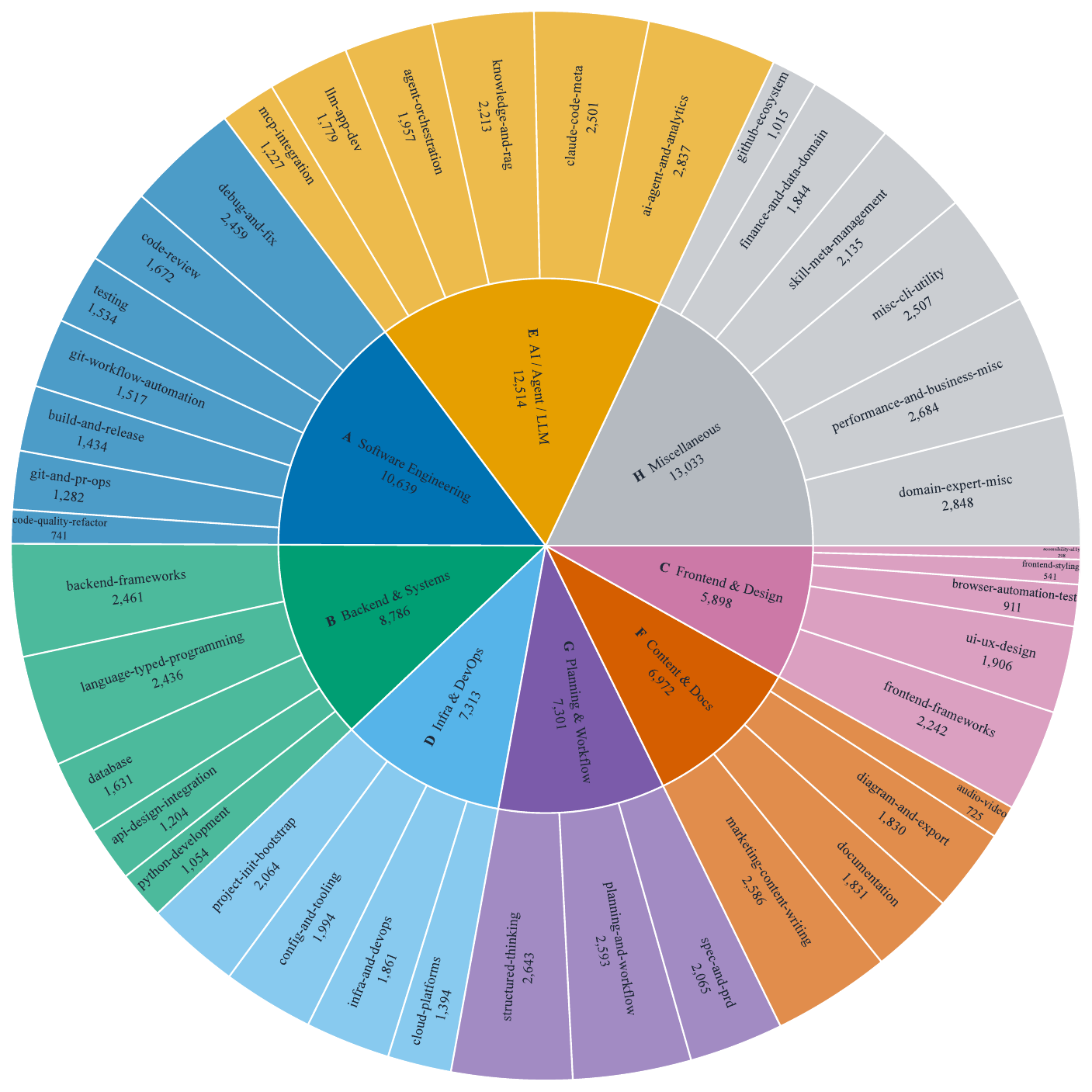}
\caption{Hierarchical taxonomy view of the R3-Skill skill pool under $K=40$ clustering. The inner ring shows the 8 super-domains (A--H), obtained by topic-merging the 40 sub-clusters; the outer ring shows the 40 sub-clusters, with sub-cluster names and arc lengths proportional to skill counts. Inner and outer rings sharing the same color family belong to the same super-domain. The figure exposes the hierarchical structure underlying the sampling rule of ``$k=2$ same sub-cluster, $k=3$ same super-domain''.}
\label{fig:cluster_projection}
\end{figure*}

\begin{figure}[htbp]
\centering
\includegraphics[width=0.85\textwidth]{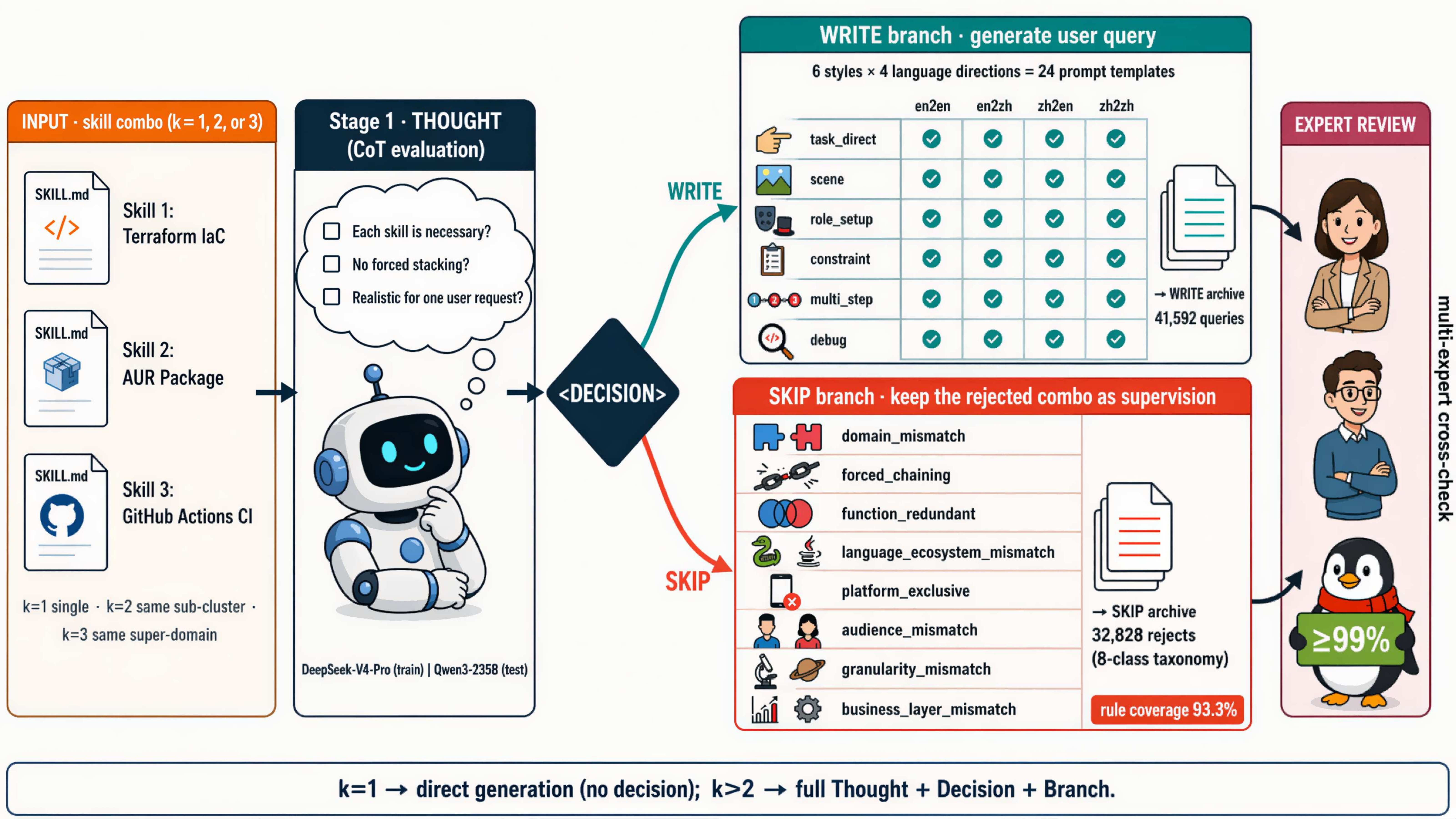}
\caption{LLM annotation flow: skill set $\to$ DECISION; the WRITE branch generates queries in 6 styles $\times$ 4 directions, while the SKIP branch is archived; both then enter taxonomy classification and multi-expert verification.}
\label{fig:annotation_flow}
\end{figure}

\begin{table}[htbp]
\centering
\caption{Query length p50 by style $\times$ direction.}
\small
\begin{tabular}{lrrrr}
\toprule
Style & en2en & en2zh & zh2zh & zh2en \\
\midrule
\emph{task\_direct} & 97 & 107 & 146 & 170 \\
\emph{scene} & 170 & 164 & 190 & 209 \\
\emph{role\_setup} & 170 & 177 & 192 & 216 \\
\emph{constraint} & 183 & 193 & 203 & 262 \\
\emph{multi\_step} & 180 & 197 & 213 & 226 \\
\emph{debug} & 194 & 176 & 196 & 227 \\
\bottomrule
\end{tabular}
\end{table}

\begin{table}[htbp]
\centering
\caption{SKIP statistics under different counting protocols.}
\small
\begin{tabular}{lrrrr}
\toprule
Protocol & k=1 & k=2 & k=3 & Total \\
\midrule
Raw SKIP & 7,163 & 15,698 & 9,967 & 32,828 \\
Classifiable SKIP & --- & 9,277 & 6,685 & 15,962 \\
Training-split SKIPs & --- & 5,978 & 3,322 & 9,300 \\
Evaluation-split SKIPs & --- & --- & --- & 16,365 \\
\bottomrule
\end{tabular}
\end{table}

\begin{table}[htbp]
\centering
\caption{Representative WRITE / SKIP examples from R3-Skill.}
\footnotesize
\setlength{\tabcolsep}{3pt}
\begin{tabular}{@{}>{\raggedright\arraybackslash}p{0.30\textwidth}c>{\raggedright\arraybackslash}p{0.28\textwidth}>{\raggedright\arraybackslash}p{0.26\textwidth}@{}}
\toprule
Skill combination & Decision & Core LLM rationale & Reject reason class \\
\midrule
Godot Multiplayer + Legal Policy Writer & SKIP & platform / domain mutual exclusion & \emph{platform\_exclusive} \\
Rust Production Reliability + Go Security Patterns & SKIP & inconsistent language ecosystems & \emph{language\_\allowbreak ecosystem\_\allowbreak mismatch} \\
Code Review + Systematic Code Review & SKIP & overlapping functionality & \emph{function\_redundant} \\
Terraform IaC + AUR Package Publish & WRITE & forms a natural release pipeline & --- \\
SendGrid Email + Xero Invoice & WRITE & forms a natural invoice-email flow & --- \\
\bottomrule
\end{tabular}
\end{table}

\begin{figure}[htbp]
\centering
\includegraphics[width=0.85\columnwidth]{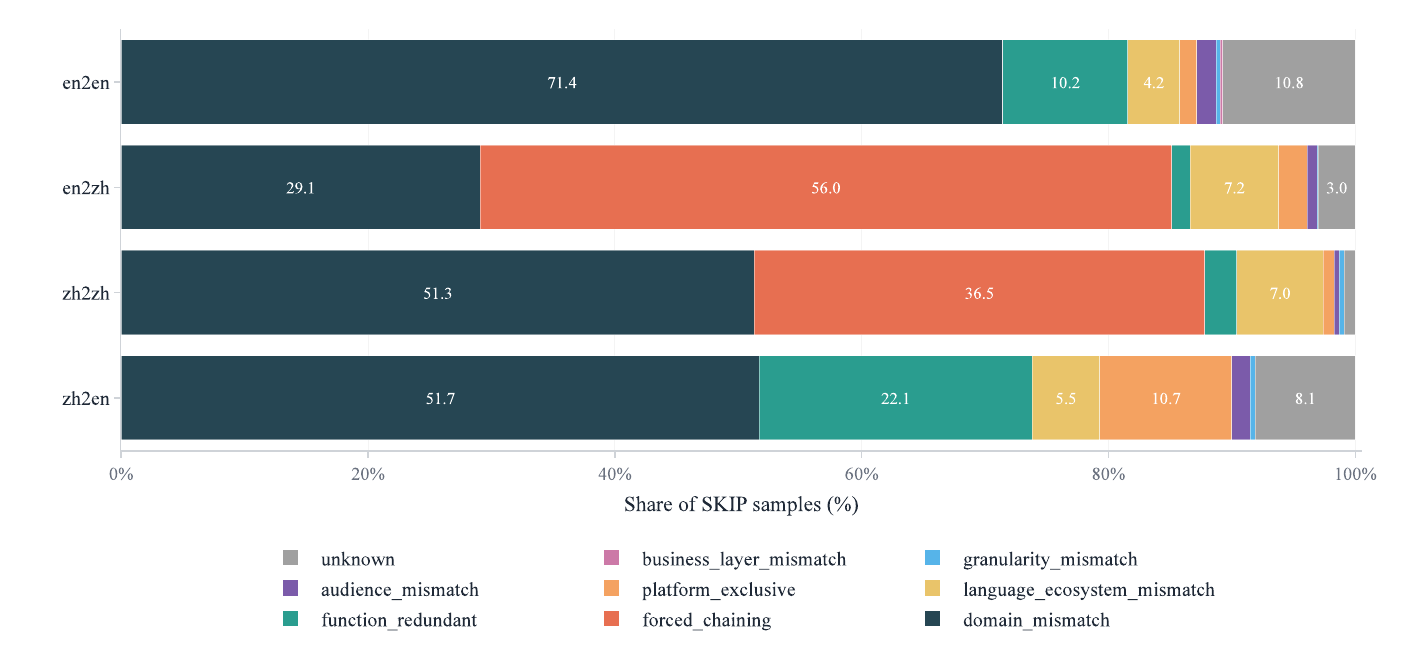}
\caption{Stacked bar chart of the 8 reject reason classes across the 4 language directions. en2en is mostly assigned to \emph{domain\_mismatch}; en2zh contains many \emph{forced\_chaining}-style rejections (0.560).}
\label{fig:taxonomy_dist}
\end{figure}

\begin{table}[htbp]
\centering
\caption{8-class reject reason taxonomy across directions and k.}
\label{tab:reject_taxonomy}
\small
\setlength{\tabcolsep}{1pt}
\begin{tabular}{@{}>{\raggedright\arraybackslash}p{0.21\textwidth}>{\raggedright\arraybackslash}p{0.18\textwidth}rrrrrrr@{}}
\toprule
Reason & Meaning & en2en & en2zh & zh2zh & zh2en & k=2 & k=3 & Share \\
\midrule
\emph{domain\_mismatch} & Domains do not overlap & 71.7\% & 29.1\% & 51.3\% & 51.7\% & 45.3\% & 46.6\% & 45.9\% \\
\emph{forced\_chaining} & Chinese ``forced stacking'' fallback & <1\% & 56.0\% & 36.5\% & <5\% & 34.6\% & 33.0\% & 33.9\% \\
\emph{function\_redundant} & Overlapping function; only one needed & 10.2\% & 1.5\% & 2.6\% & 22.1\% & 4.7\% & 4.0\% & 4.4\% \\
\emph{language\_\allowbreak ecosystem\_\allowbreak mismatch} & Different languages / ecosystems & 4.2\% & 7.2\% & 7.0\% & 5.5\% & 6.1\% & 5.1\% & 5.7\% \\
\emph{platform\_exclusive} & Platform mutual exclusion & 1.4\% & 2.3\% & 0.9\% & 10.7\% & 1.9\% & 1.9\% & 1.9\% \\
\emph{audience\_mismatch} & Different user groups & 1.6\% & 0.8\% & 0.4\% & 1.5\% & 1.4\% & 1.0\% & 1.2\% \\
\emph{granularity\_mismatch} & Low-level vs.\ high-level & 0.3\% & 0.1\% & 0.4\% & 0.4\% & 0.3\% & 0.1\% & 0.2\% \\
\emph{business\_\allowbreak layer\_\allowbreak mismatch} & Strategy vs.\ operations & 0.2\% & <0.1\% & <0.5\% & <0.5\% & 0.1\% & <0.1\% & 0.1\% \\
\emph{unknown} & No rule hit & 10.8\% & 3.0\% & 0.9\% & 8.1\% & 5.6\% & 8.8\% & 6.7\% \\
\bottomrule
\end{tabular}
\end{table}

\FloatBarrier
\section{Full R3-Skill Query Generation Prompts}
\label{app:gen-prompts}

The prompts below are included verbatim for reproducibility. Their constraints operationalize three benchmark requirements discussed in the main paper: natural user phrasing, no direct skill-name disclosure, and an explicit joint-necessity judgment for multi-skill requests.

This appendix lists all prompt templates used in the data-generation stage of the main paper (LLM Annotation). There are 8 templates in total: 4 directions (en2en / en2zh / zh2en / zh2zh; skill language then query language) $\times$ (single ($k=1$), multi ($k\geq2$, two-stage CoT WRITE/SKIP)). Each template is filled with format substitutions before use: \texttt{style\_desc} from the style table below, \texttt{k} for the skill count, and \texttt{skills\_block} for a rendered listing of skills with \emph{name} / \emph{description} / \emph{body}.

\begingroup
\small
\setlength{\parskip}{0pt}
\setlength{\partopsep}{0pt}
\setlength{\topsep}{3pt}
\setlength{\parsep}{0pt}
\makeatletter
\renewcommand\subsection{\@startsection{subsection}{2}{\z@}%
  {1.0ex}%
  {0.4ex}%
  {\normalfont\large\bfseries}}
\renewcommand\paragraph{\@startsection{paragraph}{4}{\z@}%
  {0.6ex}%
  {-1em}%
  {\normalfont\normalsize\bfseries}}
\makeatother

\Needspace{0.6\textheight}
\subsection{Style Descriptions (6 styles, bilingual)}

\prompthead{Chinese style descriptions (used by zh2zh / en2zh).}
\begin{promptbox}
\begin{Verbatim}[fontsize=\scriptsize,formatcom=\promptfont]
task_direct: 任务直述：开门见山地告诉模型要做什么，可以省略主语，像在下指令。
scene:       场景描述：先简短交代背景或痛点，再引出需求；不要使用"作为某某"的句式。
role_setup:  角色设定：让模型扮演某个专家身份再来帮你，
             例如"你是一名资深 X，请帮我..."。
constraint:  约束条件式：把目标 + 输入 + 输出格式 / 风格 / 长度 / 限制条件等讲清楚。
multi_step:  多步骤式：把需求拆成 2~4 个有先后的小步骤，让模型按顺序处理。
debug:       排错求助式：贴一段疑似有问题的代码 / 配置 / 报错，让模型定位并修复
             （也可以只描述现象）。
\end{Verbatim}
\end{promptbox}

\prompthead{English style descriptions (used by en2en / zh2en).}
\begin{promptbox}
\begin{Verbatim}[fontsize=\scriptsize,formatcom=\promptfont]
task_direct: Task-direct: state the task plainly, like an instruction.
scene:       Scene-first: briefly describe a context or pain point before stating
             the need; avoid 'As a ...' phrasing.
role_setup:  Role-setup: ask the model to act as some expert (e.g., 'You are a
             senior X, please help me ...').
constraint:  Constraint-style: spell out goal + input + output format / style /
             length / constraints.
multi_step:  Multi-step: break the request into 2~4 ordered substeps for the
             model to handle in sequence.
debug:       Debug-help: paste suspected-broken code / config / error and ask
             for diagnosis and fix (description-only is fine).
\end{Verbatim}
\end{promptbox}

\Needspace{0.55\textheight}
\subsection{Single-skill prompts ($k=1$)}

\prompthead{zh2zh --- Chinese skill $\to$ Chinese query.}
\begin{promptbox}
\begin{Verbatim}[fontsize=\scriptsize,formatcom=\promptfont]
你扮演一个真实的中文用户，正要去问一个 AI 助手帮忙做事。
你只能看到下方 skill 的能力（name / description / body 摘要），但你写出来的问题里
**绝对不能** 出现 skill 的 name 或与之高度同义的措辞，也不能照抄 description / body
的原句。

风格要求（必须严格遵循）：{style_desc}

写作硬约束：
1. 用中文写问题，长度 50~300 字；可以叠加 1~2 种次要风格让句子更自然，但主风格保持
   上面那条。
2. 不能出现 skill 的 name；不能出现明显从英文 name 直译过来的整词组合；公认缩写
   （PDF / API / SQL / GitHub 这类）允许。
3. 不要照抄 description / body 的整句话，可以重新换说法。
4. 必须能让 AI 助手读完之后判断"我应该用这个 skill 来帮他完成"。
5. 只输出问题本身，不要加任何前后缀、标签或解释。

Skill:
{skill_rendered}
\end{Verbatim}
\end{promptbox}

\prompthead{en2zh --- English skill $\to$ Chinese query.}
\begin{promptbox}
\begin{Verbatim}[fontsize=\scriptsize,formatcom=\promptfont]
你扮演一个真实的中文用户，正要去问一个 AI 助手帮忙做事。
下方给的是一份**英文 skill** 的能力（name / description / body 摘要），你需要根据它
来写一个**中文用户问题**，并且 **绝对不能** 出现 skill 的英文 name 或对它的中文直译，
也不能照抄 description / body 的原句。

风格要求（必须严格遵循）：{style_desc}

写作硬约束：
1. 用**中文**写问题，长度 50~300 字；可以叠加 1~2 种次要风格让句子更自然，但主风格
   保持上面那条。
2. 完全用中文，**不要出现英文单词**（公认缩写如 PDF / API / SQL / GitHub 这类允许）。
3. 不能出现 skill 的英文 name，也不要把它整词直译过来；可以从功能/场景去描述。
4. 不要照抄 description / body 的句子。
5. 必须能让 AI 助手判断"我应该用这个 skill"。
6. 只输出问题本身，不要加任何前后缀、标签或解释。

Skill:
{skill_rendered}
\end{Verbatim}
\end{promptbox}

\prompthead{en2en --- English skill $\to$ English query.}
\begin{promptbox}
\begin{Verbatim}[fontsize=\scriptsize,formatcom=\promptfont]
You play a real user asking an AI assistant for help.
You can only see the skill below (name / description / body excerpt). The
question you write **must not** mention the skill's name (or close synonyms),
and **must not** copy sentences from description / body verbatim.

Style requirement (strict):
{style_desc}

Hard constraints:
1. Write in English, 50~300 words. You may layer in 1~2 secondary styles to keep
   it natural, but stay anchored to the main style above.
2. Do not mention the skill's name or any near-synonym translation of it.
3. Do not copy whole sentences from description / body; rephrase.
4. The AI must be able to read your question and infer "I should use this skill".
5. Output only the question itself; no prefixes, tags, or explanations.

Skill:
{skill_rendered}
\end{Verbatim}
\end{promptbox}

\prompthead{zh2en --- Chinese skill $\to$ English query.}
\begin{promptbox}
\begin{Verbatim}[fontsize=\scriptsize,formatcom=\promptfont]
You play a real user asking an AI assistant for help in English.
The skill below is in **Chinese** (name / description / body excerpt). Write an
**English user question** based on it; you **must not** mention the skill's name
(or any direct English translation of it), and **must not** copy from description
/ body.

Style requirement (strict):
{style_desc}

Hard constraints:
1. Write in **English** only, 50~300 words. You may layer 1~2 secondary styles
   for naturalness, but the main style above must show.
2. No Chinese characters in the output.
3. Do not surface the skill's name or any direct translation of it; describe by
   function/scenario instead.
4. Do not copy sentences from description / body.
5. The AI must be able to read your question and infer "I should use this skill".
6. Output only the question itself; no prefixes, tags, or explanations.

Skill:
{skill_rendered}
\end{Verbatim}
\end{promptbox}

\Needspace{0.55\textheight}
\subsection{Multi-skill prompts ($k=2$ / $k=3$, two-stage CoT WRITE/SKIP)}

\prompthead{zh2zh --- Chinese skill set $\to$ Chinese query.}
\begin{promptbox}
\begin{Verbatim}[fontsize=\scriptsize,formatcom=\promptfont]
你扮演一个真实的中文用户，可能正在做一个比较复杂的任务，需要多个能力配合。
下面给你 {k} 个 skill 的能力（name / description / body 摘要）。请按下面两阶段输出：

第一阶段（思考）：评估这 {k} 个 skill **同时**被一个用户在 **同一句话/同一段需求**
里自然提出来的合理性。判断标准：
  - 缺一不可：去掉任何一个，需求就完成不了 / 明显残缺；
  - 不强行拼凑：不能是"我顺手再来个 X"这种硬塞；
  - 真实合理：现实中真的会有用户这么问。

第二阶段（决定）：
  - 如果合理：在 <DECISION>WRITE</DECISION> 之后输出一段中文用户问题；
  - 如果不合理：在 <DECISION>SKIP</DECISION> 之后简短说明原因，不要再写问题。

写作硬约束（仅当 WRITE 时）：
- 风格要求：{style_desc}
- 用中文写，长度 50~300 字；可叠加 1~2 种次要风格；
- 必须**自然地同时**需要这 {k} 个 skill；不要列点式"我要 A、要 B、要 C"；
- 不能出现任何一个 skill 的 name；不要照抄 description / body 原句；
- 公认缩写（PDF / API / SQL / GitHub）允许。

输出格式（**严格**）：
<THOUGHT>
你的逐条评估和最后判断，3~6 句即可。
</THOUGHT>
<DECISION>WRITE</DECISION> 或 <DECISION>SKIP</DECISION>
<QUERY>
（只在 WRITE 时填，写问题本身；SKIP 时此节留空或写一句原因）
</QUERY>

下面是 {k} 个 skill：
{skills_block}
\end{Verbatim}
\end{promptbox}

\prompthead{en2zh --- English skill set $\to$ Chinese query.}
\begin{promptbox}
\begin{Verbatim}[fontsize=\scriptsize,formatcom=\promptfont]
你扮演一个真实的中文用户，正要去问 AI 助手做一个稍复杂的事。
下面给你 {k} 个**英文 skill** 的能力（name / description / body 摘要）。先用中文
思考它们组合是否合理，再决定是否写出**中文用户问题**。

第一阶段（思考）：评估这 {k} 个 skill **同时**被一个用户在 **同一句话/同一段需求**
里自然提出来的合理性。判断标准：
  - 缺一不可：去掉任何一个，需求就完成不了 / 明显残缺；
  - 不强行拼凑；
  - 真实合理。

第二阶段（决定）：
  - 合理 -> <DECISION>WRITE</DECISION> 然后写问题；
  - 不合理 -> <DECISION>SKIP</DECISION> 简短说原因。

写作硬约束（仅当 WRITE 时）：
- 风格要求：{style_desc}
- **用中文**写，长度 50~300 字；
- 不要出现英文单词（PDF / API / SQL / GitHub 等公认缩写允许）；
- 不要出现任何一个 skill 的英文 name 或直译；
- 不要照抄 description / body 的句子。

输出格式（**严格**）：
<THOUGHT>
3~6 句中文评估。
</THOUGHT>
<DECISION>WRITE</DECISION> 或 <DECISION>SKIP</DECISION>
<QUERY>
（仅 WRITE 时填）
</QUERY>

下面是 {k} 个 skill：
{skills_block}
\end{Verbatim}
\end{promptbox}

\prompthead{en2en --- English skill set $\to$ English query.}
\begin{promptbox}
\begin{Verbatim}[fontsize=\scriptsize,formatcom=\promptfont]
You play a real user asking an AI assistant to help with a somewhat compound
task. Below are {k} skills (name / description / body excerpt). Use a two-stage
output:

Stage 1 (think): Evaluate whether these {k} skills could plausibly be needed
**together** by **one user in one request**. Tests:
  - Each is necessary: remove any and the request becomes incomplete;
  - No forced stacking ("oh and also X");
  - Realistic: a real person would phrase this.

Stage 2 (decide):
  - Plausible  -> <DECISION>WRITE</DECISION> then write the user question.
  - Not plausible -> <DECISION>SKIP</DECISION> with a one-line reason; do not
    write a question.

Writing constraints (only when WRITE):
- Style: {style_desc}
- English, 50~300 words; you may layer 1~2 secondary styles;
- Naturally require **all** {k} skills; do not enumerate "I want A, B, C";
- Do not surface any skill's name; do not copy sentences from description
  / body.

Strict output format:
<THOUGHT>
3~6 sentences of evaluation.
</THOUGHT>
<DECISION>WRITE</DECISION> or <DECISION>SKIP</DECISION>
<QUERY>
(only when WRITE)
</QUERY>

The {k} skills:
{skills_block}
\end{Verbatim}
\end{promptbox}

\prompthead{zh2en --- Chinese skill set $\to$ English query.}
\begin{promptbox}
\begin{Verbatim}[fontsize=\scriptsize,formatcom=\promptfont]
You play a real user asking an AI assistant in **English** to help with a
compound task. The {k} skills below are described **in Chinese** (name /
description / body excerpt). Reason about their joint plausibility, then decide
whether to write an **English user question**.

Stage 1 (think): Evaluate whether these {k} skills could plausibly be needed
**together** by **one user in one request**. Tests:
  - Each is necessary;
  - No forced stacking;
  - Realistic.

Stage 2 (decide):
  - Plausible  -> <DECISION>WRITE</DECISION> then write the user question
    in **English**.
  - Not plausible -> <DECISION>SKIP</DECISION> with a one-line reason.

Writing constraints (only when WRITE):
- Style: {style_desc}
- English only, 50~300 words; no Chinese characters;
- Naturally require **all** {k} skills; no enumeration;
- Do not surface any skill's name (no direct translation either);
- Do not copy sentences from description / body.

Strict output format:
<THOUGHT>
3~6 sentences of evaluation (English).
</THOUGHT>
<DECISION>WRITE</DECISION> or <DECISION>SKIP</DECISION>
<QUERY>
(only when WRITE)
</QUERY>

The {k} skills:
{skills_block}
\end{Verbatim}
\end{promptbox}
\endgroup

\end{document}